%% file: skeleton.tex
\documentclass{PoS}

\usepackage{amssymb,amsmath,amsbsy}
\usepackage{cite}
\usepackage[utf8x]{inputenc}

\usepackage{float}
\usepackage[labelfont=bf]{caption}
\usepackage{subcaption}

\usepackage{subcaption}
\usepackage{graphicx}
\usepackage{psfrag,fancyhdr,epsfig}

\input macros.tex

\title{Frame-dependence of inflationary observables in scalar-tensor gravity}

\ShortTitle{Invariants and inflationary parameters}

\author{\speaker{Alexandros Karam}\\
        Physics Department, University of Ioannina, GR--45110 Ioannina, Greece\\
        E-mail: \email{alkaram@cc.uoi.gr}}

\author{Thomas Pappas\\
        Physics Department, University of Ioannina, GR--45110 Ioannina, Greece\\
        E-mail: \email{thpap@cc.uoi.gr}}

\author{Kyriakos Tamvakis\\
        Physics Department, University of Ioannina, GR--45110 Ioannina, Greece\\
        E-mail: \email{tamvakis@uoi.gr}}

\abstract{By means of the Green's function method,
we computed the spectral indices up to third order in the slow-roll
approximation for a general scalar-tensor theory in both the Einstein and Jordan
frames. Using quantities which are invariant under the conformal rescaling of
the metric and transform as scalar functions under the reparametrization of the
scalar field, we showed that the
frames are equivalent up to this order due to the underlying assumptions.
Nevertheless, care must be taken when defining the number of $e$-folds.

}

\FullConference{Corfu Summer Institute 2018 "School and Workshops on Elementary Particle Physics and Gravity"\\
		(CORFU2018)\\
		31 August - 28 September, 2018\\
		Corfu, Greece}

\begin{document}

\section{Introduction}

The theory of cosmic inflation was originally proposed as a solution to the horizon and flatness problems~\cite{Guth1981, Linde1982}. It was soon after realized that inflation provides a mechanism to explain how quantum fluctuations in spacetime were magnified to cosmic size and became the seed for the growth of structure in the Universe~\cite{Hawking1982, Starobinsky1982, Guth1982, Linde1983a}. The simplest theory of inflation assumes the existence of one real scalar field $\phi$ which is minimally coupled to gravity, has a canonical kinetic term and is governed by a potential $V$
\begin{equation}
S = \int \dd^4 x \sqrt{- g} \left[ \frac{R}{2} - \frac{1}{2} g^{\mu\nu} \partial_\mu \phi \partial_\nu \phi - V(\phi) \right] \,, \qquad (M^2_{\rm Pl} \equiv 1) \,.
\end{equation}
The energy density of this field acts as a cosmological constant and leads to the exponential expansion of the Universe during the inflationary epoch. The Friedmann equations which describe the expansion of the Universe are
\begin{equation}
\left( \frac{\dot{a}}{a} \right)^2 \equiv H^2 = \frac{1}{3} \left[ \frac{\dot{\phi}^2}{2} + V \right] \quad
\dot{H} = - \frac{1}{2} \dot{\phi}^2 \,,
\end{equation}
where $H$ is the Hubble parameter, $a$ is the scale factor and a dot denotes differentiation with respect to cosmic time $t$. The Klein-Gordon equation which describes the dynamics of the inflaton field has the form
\begin{equation}
\ddot{\phi} + 3 H \dot{\phi} + V' = 0 \,,
\end{equation}
where a prime denotes differentiation with respect to the field $\phi$.

Inflation is usually studied with the help of the so-called slow-roll approximation. In this approximation, the field slowly rolls towards the minimum of the potential so that its potential energy dominates over its kinetic $V(\phi) \gg \dot{\phi}^2$, while the condition $\vert \ddot{\phi} \vert \ll \vert 3 H \dot{\phi} \vert , \, \vert V' \vert$ must also hold in order for inflation to last long enough to solve the horizon and flatness problems.

The first Hubble slow-roll parameter (HSRP) is defined as
\begin{equation}
\epsilon_H = - \frac{\dot{H}}{H^2} = \frac{3 \dot{\phi}^2}{\dot{\phi}^2 + 2 V} \,,
\end{equation}
and is small during inflation but becomes \textbf{exactly} unity when the acceleration of the scale factor goes to zero and inflation ends
\begin{equation}
\frac{\ddot{a}}{a} = H^2 (1 - \epsilon_H) \,.
\end{equation}
The second HSRP is defined as
\begin{equation}
\eta_H = - \frac{\ddot{\phi}}{H \dot{\phi}} \,,
\end{equation}
and is related to the condition $\vert \ddot{\phi} \vert \ll \vert 3 H \dot{\phi} \vert , \, \vert V' \vert$. In the slow-roll approximation, the equations of motion take a simpler form
\begin{equation}
H^2 \approx  \frac{1}{3} V(\phi)\, ,\qquad \dot{\phi} \approx  - \frac{V'}{3 H} \,.
\end{equation}

The shape of the potential is encoded in the so-called potential slow-roll parameters (PSRPs)
\begin{equation}
\epsilon_V = \frac{1}{2} \left( \frac{V'}{V} \right)^2 \, , \qquad  \eta_V = \frac{V''}{V} \,.
\end{equation}
The PSRPs are related to the Hubble ones through the following equations~\cite{Liddle1994}
\begin{equation}
\epsilon_V = \epsilon_H \left( \frac{3 - \eta_H}{3 - \epsilon_H} \right)^2  \, , \qquad
\eta_V = \sqrt{2 \epsilon_H} \frac{\eta'_H}{3 - \epsilon_H} + \left( \frac{3 - \eta_H}{3 - \epsilon_H} \right)  \left( \epsilon_H + \eta_H \right) \,,
\end{equation}
and if we employ a Taylor expansion we find
\begin{equation}
\epsilon_H \simeq  \epsilon_V \,, \qquad
\eta_H \simeq  \eta_V - \epsilon_V \,,
\end{equation}
to first order. The quantities that are most relevant for inflationary model building and constrained by the experiments are the scalar spectral index $n_s$ and the tensor-to-scalar ratio $r$, which we ideally want to express in terms of the PSRPs
\begin{equation}
n_s = 1 - 4 \epsilon_H + 2 \eta_H \simeq 1 - 6 \epsilon_V + 2 \eta_V  \, , \qquad
r = 16 \epsilon_H \simeq 16 \epsilon_V \,,
\end{equation}
since when we construct an inflationary model we usually consider a specific potential. Finally, the variation of the field during inflation can be connected to the number of $e$-folds
\begin{equation}
N(\phi) = \int^{t_{\mathrm{end}}}_t H \dd t = \int^\phi_{\phi_{\mathrm{end}}} \frac{\dd \phi}{\sqrt{2 \epsilon_H}} \approx \int^\phi_{\phi_{\mathrm{end}}} \frac{\dd \phi}{\sqrt{2 \epsilon_V}} \sim 50-60 \,,
\end{equation}
which give the exponential variation of the scale factor and have to be between $50$ and $60$ so that the horizon and flatness problems are solved.

The simplest models like $\phi^4$, $\phi^3$ and $\phi^2$ have already been ruled out by the Planck $2015$ data~\cite{Ade2015}, while a little more convoluted models, such as the $\alpha$-attractors~\cite{Kallosh2013b}, the Starobinsky~\cite{Starobinsky1980} and some non-minimally coupled models such as Higgs inflation~\cite{Bezrukov2008, Bezrukov2009} yield predictions that still comply with the observations.

\section{Invariant formalism and slow-roll approximation}

Most of these models belong to the general class of scalar-tensor theories~\cite{Faraoni2004a}
\begin{equation}
S = \int \mathrm{d}^4 x \sqrt{-g} \left\lbrace \frac{1}{2}\mathcal{A}(\Phi) R - \frac{1}{2}\mathcal{B}(\Phi) g^{\mu\nu} \left( \nabla_\mu \Phi \right) \left( \nabla_\nu \Phi \right) - \mathcal{V}(\Phi) \right\rbrace + S_m \left[ e^{2 \sigma(\Phi)} g_{\mu\nu} , \chi \right] \,,
\label{eq:action}
\end{equation}
where the function $\mathcal{A}$ gives the coupling to curvature, the function $\mathcal{B}$ is a general coupling with the kinetic term, $\mathcal{V}$ is the scalar potential and $\sigma$ is the conformal coupling between the metric $g_{\mu\nu}$ and the matter fields. By choosing the form of these functions we get a specific model. A conformal metric rescaling and field redefinition can fix two of the model functions to get different parametrizations, e.g.,
\begin{itemize}
\item Jordan frame Boisseau-Esposito-Far\`{e}se-Polarski-Starobinski parametrization
\[
\mathcal{A} = F(\phi), \quad \mathcal{B} = 1 , \quad \mathcal{V} = \mathcal{V} (\phi) , \quad \sigma = 0 ,
\]
\item Jordan frame Brans-Dicke-Bergmann-Wagoner parametrization
\[
\mathcal{A} = \Psi, \quad \mathcal{B} = \frac{\omega(\Psi)}{\Psi} , \quad \mathcal{V} = \mathcal{V} (\Psi) , \quad \sigma = 0  ,
\]
\item Einstein frame canonical parametrization
\[
\mathcal{A} = 1, \quad \mathcal{B} = 2 , \quad \mathcal{V} = \mathcal{V} (\varphi) , \quad \sigma = \sigma(\varphi)  .
\]
\end{itemize}
Of course, in this kind of theories the frame issue arises: which frame is physical? Jordan or Einstein? Or both? In the Jordan frame we have in general a non-minimal coupling between the scalar field and gravity but freely-falling objects made of matter follow geodesics of the metric, while in the Einstein frame we have a minimal coupling to gravity but the coupling of the matter fields to the metric are rescaled by the conformal factor. The two frames are mathematically equivalent at the classical level\footnote{See also \cite{Kamenshchik2015, Herrero-Valea2016, Pandey2016, Pandey2017, Ruf2018} for considerations on the quantum equivalence of the frames.} since one can always switch between them by applying a conformal transformation of the metric and a field redefinition, collectively referred to as \textit{frame transformation}. Nevertheless, the physical equivalence of the frames with respect to the physical predictions has become a matter of a long-standing debate \cite{Capozziello1997, Dick1998, Faraoni1999, Faraoni1999a, Flanagan2004, BHADRA2007, Nozari2009, Capozziello2010, Corda2011, Qiu2012, Qiu2012a, Quiros2013, Chiba2013a, Postma2014, Qiu2015, Domenech2015a, Burns2016, Bahamonde2016, Brooker2016, Karamitsos2017, Bhattacharya2017a, Bahamonde2017, Karamitsos2018, Bhattacharya2018, Karam2018, Quiros2018, Chakraborty2019, Falls2018, Quiros2019}.

So, in~\cite{Karam2017}, in order to avoid the frame issue we consider quantities that are invariant under a conformal rescaling of the metric and a scalar field redefinition. We write the equations of motion in terms of these invariants, then we compute the inflationary observables up to third order in the slow-roll approximation and express them in terms of the invariants.

Now, the action~\eqref{eq:action} remains invariant under a conformal transformation and field redefinition
\begin{equation}
g_{\mu\nu} = e^{2 \bar{\gamma} (\bar{\Phi})} \bar{g}_{\mu\nu} \,, \qquad  \Phi = \bar{f} (\bar{\Phi}) \,,
\end{equation}
if the model functions transform as~\cite{Flanagan2004}
\begin{eqnarray}
\bar{\mathcal{A}}(\bar{\Phi}) &=& e^{2 \bar{\gamma} (\bar{\Phi})} \mathcal{A} \left( \bar{f} (\bar{\Phi}) \right) , \\
\bar{\mathcal{B}}(\bar{\Phi}) &=& e^{2 \bar{\gamma} (\bar{\Phi})} \left[ (\bar{f}')^2 \mathcal{B} \left( \bar{f} (\bar{\Phi}) \right) - 6 (\bar{\gamma}')^2 \mathcal{A} \left( \bar{f} (\bar{\Phi}) \right) - 6 \bar{\gamma}' \bar{f}' \mathcal{A}'  \right] , \\ 
\bar{\mathcal{V}} (\bar{\Phi}) &=& e^{4 \bar{\gamma} (\bar{\Phi})} \mathcal{V} \left( \bar{f} (\bar{\Phi}) \right) , \\
\bar{\sigma} (\bar{\Phi}) &=& \sigma \left( \bar{f} (\bar{\Phi})  \right) + \bar{\gamma} (\bar{\Phi}) , 
\end{eqnarray}
By using these transformation rules we can construct combinations of the model functions that remain invariant under a frame transformation.

The scalar invariants that we use are~\cite{Jaerv2015}
\begin{eqnarray}
\Inm (\Phi) &\equiv & \frac{e^{2 \sigma (\Phi)}}{\mathcal{A} (\Phi)}  \\
\Iv (\Phi) &\equiv & \frac{\mathcal{V} (\Phi)}{(\mathcal{A} (\Phi))^2} \label{inv_v}  \\
\Iphi (\Phi) &\equiv & \int \left( \frac{2 \mathcal{A} \mathcal{B} + 3 (\mathcal{A}')^2}{4 \mathcal{A}^2} \right)^{1/2} \mathrm{d} \Phi 
\end{eqnarray}
The first one is related to the non-minimal coupling, the second one plays the role of the invariant potential, while the third one can be interpreted as the invariant field.

\begin{figure}
\includegraphics[width=.5\textwidth]{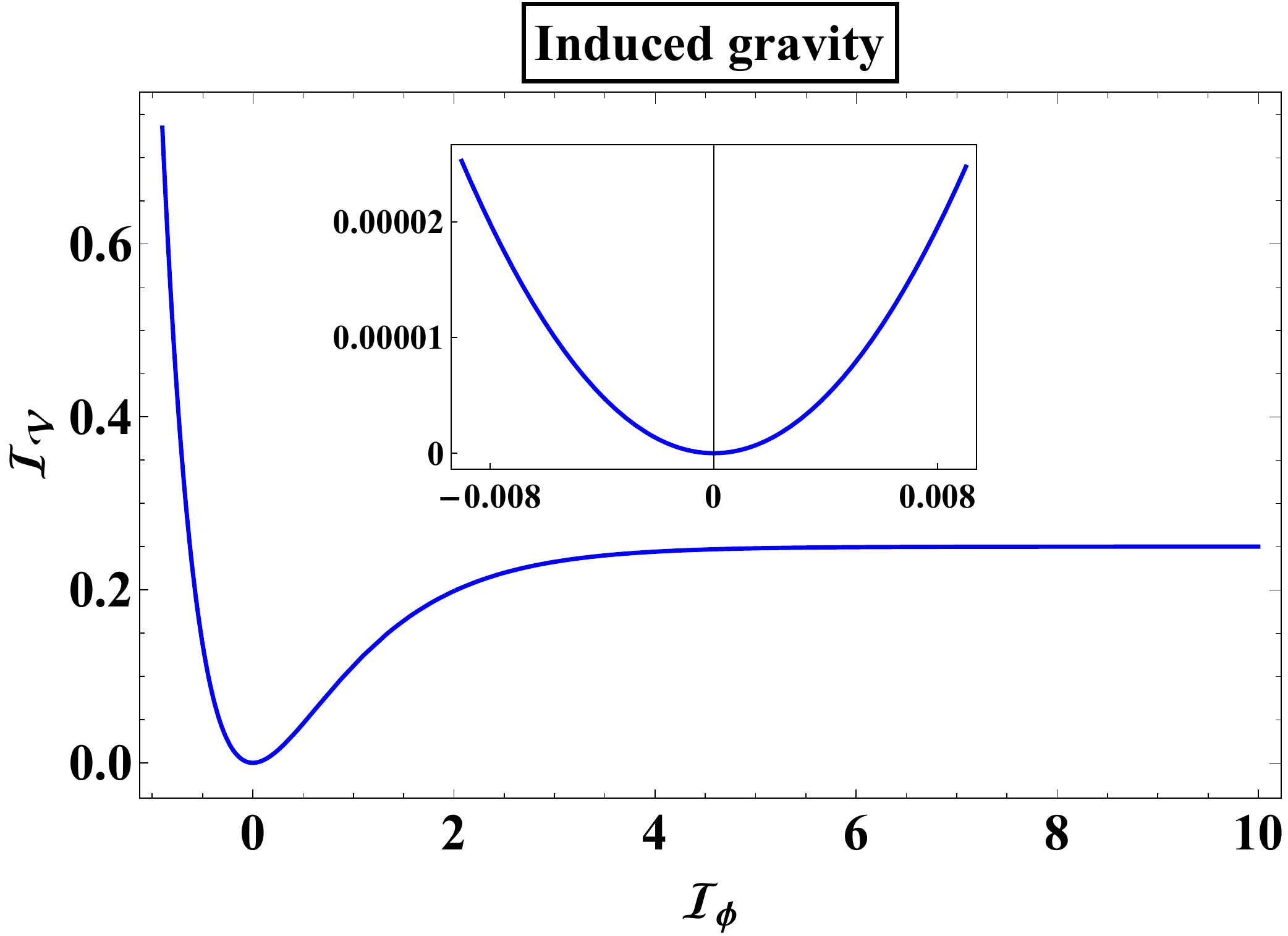}
\includegraphics[width=.5\textwidth]{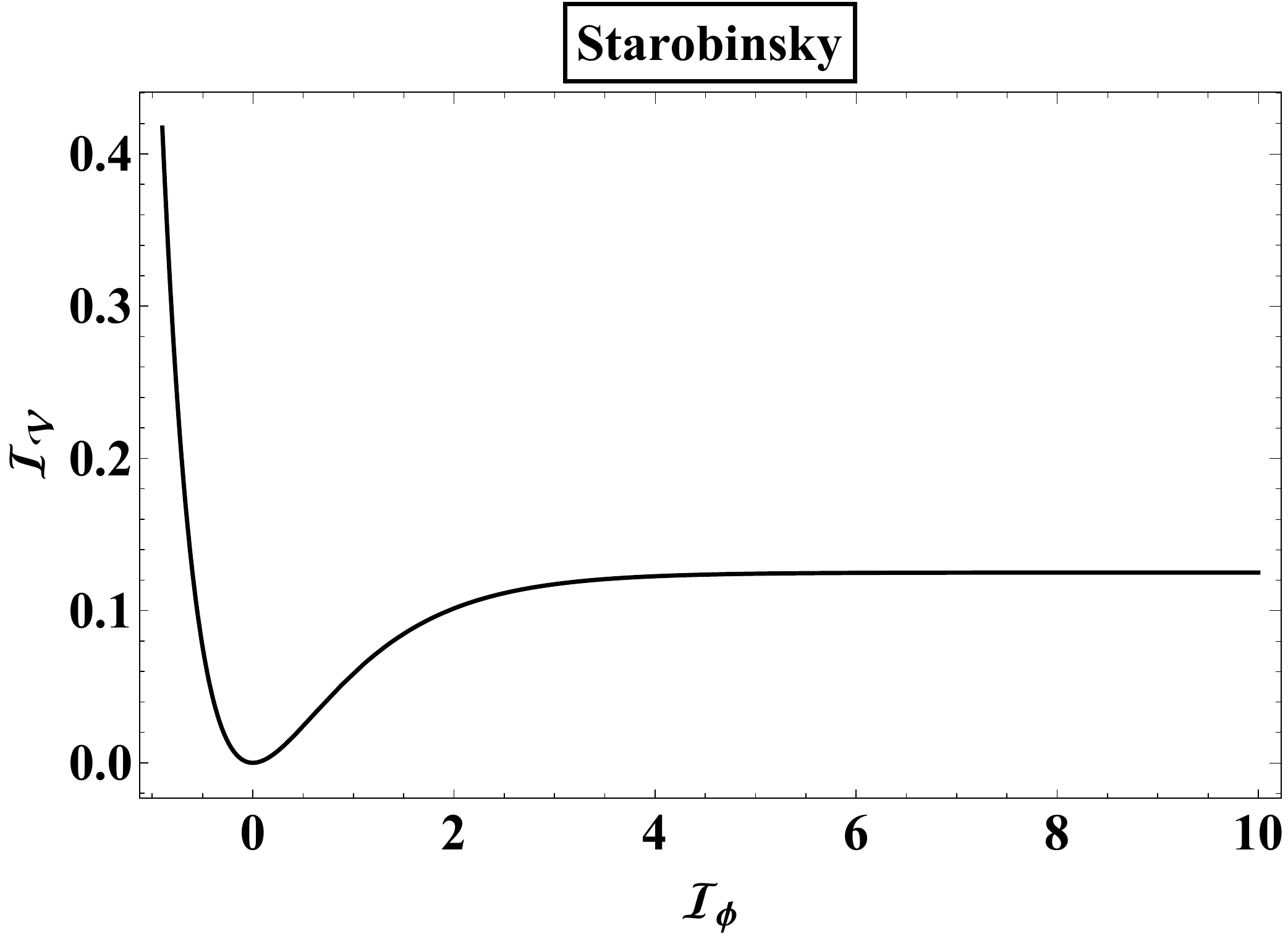}
\caption{The normalized invariant inflationary potentials for induced gravity and Starobinsky models for $\xi=2$. In the strong coupling limit the invariant potentials have a similar form and lead to the same predictions, while in the weakly-coupled limit induced gravity approaches the quadratic inflation attractor (inset in left plot).}
\label{fig:Induced-v-Staro}
\end{figure}

A very attractive feature of the invariant formalism is that we can easily classify inflationary models based on their invariant potential~\cite{Jaerv2017}. For example, let us consider \textit{induced gravity} inflation \cite{Accetta1985} and \textit{Starobinsky} inflation \cite{Starobinsky1980}. The former is described by the model functions
\begin{equation}
\mathcal{A} (\Phi) = \xi \Phi^2 \,, \quad \mathcal{B} (\Phi) = 1 \,, \quad \sigma (\Phi) = 0 \,, \quad \mathcal{V} (\Phi) = \lambda \left( \Phi^2 - v^2 \right)^2\,
\end{equation}
where $\xi$ is the nonminimal coupling and $v$ is the vacuum expectation value (VEV) of the scalar field $\Phi$ which induces the Planck mass scale, $1 = \xi v^2$.
For Starobinsky inflation with $f(R) = R + b R^2$ one has~\cite{Burns2016}
\begin{equation}
\mathcal{A} (\Phi) = \Phi \,, \quad \mathcal{B} (\Phi) = 0 \,, \quad \sigma (\Phi) = 0 \,, \quad \mathcal{V} (\Phi) = \frac{b}{2} \left( \frac{\Phi-1}{2b}\right)^2 \,.
\end{equation}
From the above we calculate the form of the invariant fields, invert them to find $\Phi(\Iphi)$ and then using \eqref{inv_v} we calculate $\Iv(\Phi(\Iphi)) = \Iv(\Iphi)$ and obtain
\ba
\text{Induced gravity:}&& \qquad \Iv (\Iphi)= \frac{\lam}{\xi^2}\left( 1- e^{-\sqrt{\frac{8 \xi}{1+6 \xi}}\Iphi} \right)^2,
\label{Induced_Inv_pot}
\\
\text{Starobinsky:}&& \qquad \Iv (\Iphi)=\frac{1}{8 \, b} \left( 1- e^{-\frac{2}{\sqrt{3}}\Iphi}\right)^2.
\label{Staro_Inv_pot}
\ea
For relatively large values of the non-minimal coupling they have the same form, see Fig.~\ref{fig:Induced-v-Staro}.
This explains why such different models give essentially the same predictions for the inflationary observables.

Apart from scalar invariants we can also define tensorial invariants. By using the transformation properties of the model functions we define the invariant metric
\begin{equation}
\hat{g}_{\mu\nu} \equiv \mathcal{A}(\Phi) g_{\mu\nu}
\end{equation}
Note that this is not a unique choice. For example, by multiplying $\hat{g}$ with the first invariant we the $\bar{g}$ metric
\begin{equation}
\bar{g}_{\mu\nu} \equiv e^{2 \sigma (\Phi)} g_{\mu\nu} = \Inm \hat{g}_{\mu\nu}
\end{equation}
which is also invariant under conformal transformations and field redefinitions. Notice that matter fields couple to the $\bar{g}$ metric so we call it the invariant Jordan frame metric, while we call $\hat{g}$ the invariant Einstein frame metric. Assuming a flat FLRW spacetime, 
\begin{equation}
\mathrm{d}s^2 \equiv g_{\mu\nu}\mathrm{d}x^\mu \mathrm{d}x^\nu = -\mathrm{d}t^2 + \left(a(t)\right)^2 \delta_{ij}\mathrm{d}x^i \mathrm{d}x^j
\end{equation}
because of homogeneity and isotropy the inflaton and the invariants only depend on the cosmic time $t$, i.e.\ $\mathcal{I}_{i} = \mathcal{I}_{i}(t)$ etc. From the FLRW metric we can obtain the time parameter and scale factor in the invariant Einstein frame
\begin{equation}
\frac{\mathrm{d} \phantom{\hat{t}}}{\mathrm{d}\hat{t}} \equiv \frac{1}{\sqrt{\mathcal{A}}}\frac{\mathrm{d} \phantom{{t}}}{\mathrm{d}t} \,, \qquad
\hat{a}(\hat{t} ) \equiv \sqrt{\mathcal{A}} \, a(t)
\end{equation}
and from there to the invariant Jordan frame
\begin{equation}
\frac{\mathrm{d} \phantom{\hat{t}}}{\mathrm{d}\bar{t}} = \frac{1}{\sqrt{\Inm}}\frac{\mathrm{d} \phantom{{t}}}{\mathrm{d}\hat{t}} \,, \quad
\bar{a}(\bar{t} )  =\sqrt{\Inm}\hat{a}(\hat{t} )\,, \quad
\bH \equiv \frac{1}{\bar{a}} \frac{\mathrm{d} \bar{a} }{\mathrm{d}\bar{t}}= \frac{1}{\sqrt{\Inm}} \left( \hH + \frac{1}{2} \frac{\mathrm{d} \ln{\Inm}}{\mathrm{d} \hatt} \right)
\end{equation}

The equations of motion in terms of the invariants in the invariant Einstein frame have the form
\begin{equation}
\hH^2 = \frac{1}{3} \left[ \left( \frac{\mathrm{d} \Iphi}{\mathrm{d} \hatt} \right)^2 + \Iv \right] ,
\quad 
\frac{\mathrm{d} \hH}{\mathrm{d} \hatt} = - \left( \frac{\mathrm{d} \Iphi}{\mathrm{d} \hatt} \right)^2 ,
\quad 
\frac{\mathrm{d}^2 \Iphi}{\mathrm{d} \hatt^2} = - 3 \hH \frac{\mathrm{d} \Iphi}{\mathrm{d} \hatt} - \frac{1}{2} \frac{\mathrm{d} \Iv}{\mathrm{d} \Iphi}
\end{equation}
and the first two HSRPs are defined as
\begin{equation}
\heps_0 \equiv - \frac{1}{\hH^2} \frac{\mathrm{d} \hH}{\mathrm{d} \hatt}  , \qquad  \hat{\eta} \equiv - \left( \hH \frac{\mathrm{d} \Iphi}{\mathrm{d} \hatt} \right)^{-1} \frac{\mathrm{d}^2 \Iphi}{\mathrm{d} \hatt^2}
\end{equation}
Also, since we wish to compute the inflationary observables to higher-order in the slow-roll approximation, we introduce the following series of parameters:
\begin{equation}
\hk_0 \equiv \frac{1}{\hH^2} \left( \frac{\mathrm{d} \Iphi}{\mathrm{d} \hatt} \right)^2  ,
\quad 
\hk_1 \equiv \frac{1}{\hH \hk_0} \frac{\mathrm{d} \hk_0}{\mathrm{d} \hatt} = 2 \left( - \hat{\eta} + \heps_0 \right) ,
\quad
\hk_{i+1} \equiv \frac{1}{\hH \hk_i} \frac{\mathrm{d} \hk_i}{\mathrm{d} \hatt} 
\end{equation}
Now, in the invariant Jordan frame the equations of motion have the form
\begin{eqnarray}
\bH^2 &=& \frac{1}{3} \left( \frac{\mathrm{d} \Iphi}{\mathrm{d} \bart} \right)^2 + \bH \frac{\mathrm{d} \ln{\Inm}}{\mathrm{d} \bart} - \frac{1}{4} \left( \frac{\mathrm{d} \ln{\Inm}}{\mathrm{d} \bart} \right)^2 + \frac{1}{3} \frac{\Iv}{\Inm} \\
\frac{\mathrm{d}^2 \Iphi}{\mathrm{d} \bart^2} &=& \left( - 3 \bH + \frac{\mathrm{d} \ln{\Inm}}{\mathrm{d} \bart} \right) \frac{\mathrm{d} \Iphi}{\mathrm{d} \bart} - \frac{1}{2 \Inm} \frac{\mathrm{d} \Iv}{\mathrm{d} \Iphi} \\
\frac{\mathrm{d} \bH}{\mathrm{d} \bart} &=& - \frac{1}{2} \bH \frac{\mathrm{d} \ln{\Inm}}{\mathrm{d} \bart} + \frac{1}{4} \left( \frac{\mathrm{d} \ln{\Inm}}{\mathrm{d} \bart} \right)^2 - \left( \frac{\mathrm{d} \Iphi}{\mathrm{d} \bart} \right)^2 + \frac{1}{2} \frac{\mathrm{d}^2 \ln{\Inm}}{\mathrm{d} \bart^2}
\end{eqnarray}
and the standard HSRPs have the form
\begin{equation}
\bar{\epsilon}_0 \equiv - \frac{1}{\bH^2} \frac{\mathrm{d} \bH}{\mathrm{d} \bart} , \qquad \bar{\eta} \equiv - \left( \bH \frac{\mathrm{d} \Iphi}{\mathrm{d} \bart} \right)^{-1} \frac{\mathrm{d}^2 \Iphi}{\mathrm{d} \bart^2}
\end{equation}
At this point, apart from the $\kappa$s
\begin{equation}
\bk_0 \equiv \frac{1}{\bH^2} \left( \frac{\mathrm{d} \Iphi}{\mathrm{d} \bart} \right)^2 ,
\quad 
\bk_1 \equiv \frac{1}{\bH \bk_0} \frac{\mathrm{d} \bk_0}{\mathrm{d} \bart} = 2 \left( - \bar{\eta} + \bar{\epsilon}_0 \right) ,
\quad
\bk_{i+1} \equiv \frac{1}{\bH \bk_i} \frac{\mathrm{d} \bk_i}{\mathrm{d} \bart}  
\end{equation}
we introduce a new series of HSRPs, namely,
\begin{equation}
\bl_0 \equiv \frac{1}{2 \bH} \frac{\mathrm{d} \ln{\Inm}}{\mathrm{d} \bart} ,
\quad 
\bl_1 \equiv \frac{1}{\bH \bl_0} \frac{\mathrm{d} \bl_0}{\mathrm{d} \bart} ,
\quad
\bl_{i+1} \equiv \frac{1}{\bH \bl_i} \frac{\mathrm{d} \bl_i}{\mathrm{d} \bart} 
\end{equation}
that depend on the first invariant.

Interestingly, the PSRPs depend only on invariants. The first one assumes the form~\cite{Kuusk2016}
\begin{equation}
\eps_V=\frac{1}{4 \Iv^2}\left(\frac{\dd \Iv}{\dd \Iphi} \right)^2 ,
\label{eps_V}
\end{equation}
while $\eta_V$ and higher-order parameters can be encoded in the hierarchy
\begin{equation}
{}^n\beta_V \equiv \left( \frac{1}{2 \Iv} \right)^{n}\left( \frac{\dd \Iv}{\dd \Iphi} \right)^{n-1} \left( \frac{\dd^{(n+1)} \Iv}{\dd \Iphi^{(n+1)}} \right) ,
\label{hier_V}
\end{equation}
where ${}^n\beta_V$ is a parameter of order $n$ in the slow-roll approximation. The first three parameters arising from this hierarchy are
\begin{eqnarray}
\eta_V &=& \frac{1}{2 \Iv}\left( \frac{\dd^2 \Iv}{\dd \Iphi^2}\right) ,
\label{eta_V}
\\
\zeta_V^2 &=& \frac{1}{4 \Iv^2}\left(\frac{\dd \Iv}{\dd \Iphi} \right) \left(\frac{\dd^3 \Iv}{\dd \Iphi^3} \right) ,
\label{zeta_V}
\\
\rho_V^3 &=& \frac{1}{8 \Iv^3}\left( \frac{\dd^2 \Iv}{\dd \Iphi^2}\right) \left( \frac{\dd^4 \Iv}{\dd \Iphi^4}\right) .
\label{rho_V}
\end{eqnarray}
Note that we have changed the symbols $\xi$ and $\sigma$ of \cite{Liddle1994} in order to avoid confusion with the nonminimal coupling and one of the model functions, respectively.

\section{Higher-order spectral indices}

Since we wanted to compute the inflationary observables to third order in the slow-roll approximation, we saw that the conventional methods based on the Hankel functions fail because of the underlying assumption that the first slow-roll parameters are constant during the superhorizon evolution of the curvature perturbations. Instead, we employed the Green's function method of Stewart and Gong~\cite{Gong2001}\footnote{See \cite{Stewart2002, Gong2004, Kim2004, Wei2004, Kim2005, Kadota2005, Joy2005, Dvorkin2011, Adshead2011, Kumazaki2011, Miranda2012, Adshead2013, BeltranJimenez2013, Adshead2014, Gong2014, Achucarro2014, Motohashi2015, Motohashi2017} for various extensions and applications of this method} which is valid to all orders in the slow-roll expansion. Thus, in the Einstein frame and in terms of the HSRPs, we obtained the scalar power spectrum~\cite{Karam2017}
\be 
\begin{split}
\hat{P}_{S} =
\left[ \frac{\hat{H}^4}{2(2\pi)^2} \left(\frac{\dd  \Iphi }{\dd \hat{t}}\right)^{-2}  \right]& \left[  1+(2\alp-2)\hat{\kap}_0+\alp
\hat{\kap}_1+\left( 2\alp^2-2\alp-5+\frac{\pi^2}{2} \right)\hat{\kap}_0^2
\right. \\
  &\quad \left. {}
+\left( \frac{\alp^2}{2}-1+\frac{\pi^2}{8}\right)\hat{\kap}_1^2+\left( \alp^2+\alp-7+\frac{7 \pi^2}{12} \right)\hat{\kap}_0\hat{\kap}_1
\right. \\
  &\quad \left. {}
+\left( -\frac{\alp^2}{2}+\frac{\pi^2}{24} \right)\hat{\kap}_1\hat{\kap}_2 \right] ,
\label{scalar-spectrum-EF}
\end{split}
\ee
where $ \alp  \equiv (2-\ln{2}-\gamma) \simeq 0.729637$ and $ \gamma \simeq 0.577216$, the scalar spectral index
\be 
\begin{split}
\hat{n}_{S} =&
1-2 \hat{\kap}_0-\hat{\kap}_1-2\hat{\kap}_0^2+\alp\hat{\kap}_1\hat{\kap}_2+(2\alp-3)\hat{\kap}_0 \hat{\kap}_1-2\hat{\kap}_0^3+(6\alp-17+\pi^2)\hat{\kap}_0^2\hat{\kap}_1 
\\
& +\left(-2+ \frac{\pi^2}{4} \right)\hat{\kap}_1^2\hat{\kap}_2+\left( -\frac{\alp^2}{2} +\frac{\pi^2}{24}\right)\hat{\kap}_1\hat{\kap}_2^2+\left( -\alp^2+3 \alp-7+\frac{7 \pi^2}{12} \right)\hat{\kap}_0\hat{\kap}_1^2
\\
& 
+ \left(-\frac{\alp^2}{2}+ \frac{\pi^2}{24} \right)\hat{\kap}_1\hat{\kap}_2\hat{\kap}_3+\left( -\alp^2+4\alp-7+\frac{7 \pi^2}{12} \right)\hat{\kap}_0\hat{\kap}_1\hat{\kap}_2 ,
\end{split}
\label{scalar-index-EF}
\ee
and finally the tensor-to-scalar ratio
\be
\hat{r} = 16 \hat{\kap}_0 \left[ 1-\alp\hat{\kap}_1 +\left(-\alp+5-\frac{\pi^2}{2} \right)\hat{\kap}_0\hat{\kap}_1+\left( \frac{\alp^2}{2}+1-\frac{\pi^2}{8}\right)\hat{\kap}_1^2
+\left( \frac{\alp^2}{2}-\frac{\pi^2}{24} \right)\hat{\kap}_1\hat{\kap}_2  \right] .
\label{ratio-EF}
\ee
Similarly, in the Jordan frame the second-order-corrected scalar power spectrum in the slow-roll approximation is:
\be 
\begin{split}
\bar{P}_{S} =
\left[ \frac{\bar{H}^4}{(2\pi)^2}\frac{\Inm}{2} \left(\frac{\dd  \Iphi }{\dd\bar{t}}\right)^{-2}  \right]& \left[  1-4\bar{\lam}_0+(2\alp-2)\bar{\kap}_0+\alp\bar{\kap}_1+\left(2\alp^2-2 \alp-5+\frac{\pi^2}{2} \right)\bar{\kap}_0^2
\right. \\
  &\quad \left. {}
+(4-4\alp)\bar{\lam}_0\bar{\kap}_0+(-3\alp)\bar{\lam}_0\bar{\kap}_1+\left( \frac{\alp^2}{2}-1+\frac{\pi^2}{8}\right)\bar{\kap}_1^2+6\bar{\lam}_0^2
\right. \\
  &\quad \left. {}
+2\bar{\alp}\bar{\lam}_0\bar{\lam}_1+\left( \alp^2+\alp-7+\frac{7\pi^2}{12} \right)\bar{\kap}_0\bar{\kap}_1+ \left( -\frac{\alp^2}{2}+\frac{\pi^2}{24} \right)\bar{\kap}_1\bar{\kap}_2 \right].
\label{scalar-spectrum-JF}
\end{split}
\ee
and the scalar spectral index is:
\be 
\begin{split}
\bar{n}_{S} =&
1-2 \bar{\kap}_0-\bar{\kap}_1-2\bar{\kap}_0^2-2\bar{\lam}_0\bar{\lam}_1+\alp\bar{\kap}_1\bar{\kap}_2-\bar{\kap}_1\bar{\lam}_0-4\bar{\kap}_0\bar{\lam}_0+(2\alp-3)\bar{\kap}_1 \bar{\kap}_0-2\bar{\kap}_0^3-8\bar{\lam}_0\bar{\kap}_0^2
 \\  
& -6\bar{\lam}_0^2\bar{\kap}_0+(6\alp-17+\pi^2)\bar{\kap}_0^2\bar{\kap}_1-\bar{\kap}_1\bar{\lam}_0^2+\left( -2+\frac{\pi^2}{4} \right)\bar{\kap}_1^2\bar{\kap}_2-4\bar{\lam}_0^2\bar{\lam}_1+2\alp\bar{\lam}_0 \bar{\lam}_1^2
\\
& +\left(-\frac{\alp^2}{2}+ \frac{\pi^2}{24} \right)\bar{\kap}_1\bar{\kap}_2^2+\left( -\alp^2+3 \alp-7+\frac{7 \pi^2}{12} \right)\bar{\kap}_0\bar{\kap}_1^2 +2\alp\bar{\lam}_0\bar{\lam}_1\bar{\lam}_2+(6\alp-9)\bar{\lam}_0\bar{\kap}_0\bar{\kap}_1
\\
& +(4\alp-4)\bar{\lam}_0\bar{\lam}_1\bar{\kap}_0+(\alp+1)\bar{\kap}_1\bar{\lam}_0\bar{\lam}_1+2\alp\bar{\lam}_0\bar{\kap}_1\bar{\kap}_2
+ \left( -\frac{\alp^2}{2}+\frac{\pi^2}{24} \right)\bar{\kap}_1\bar{\kap}_2\bar{\kap}_3
\\
& +\left(-\alp^2+ 4\alp-7+\frac{7 \pi^2}{12} \right)\bar{\kap}_0\bar{\kap}_1\bar{\kap}_2 .
\end{split}
\label{scalar-index-JF}
\ee
Finally, for the tensor-to-scalar ratio we have
\be
\begin{split}
\bar{r} = & 16 \bar{\kap}_0 \left[ 1+2\bar{\lam}_0-\alp\bar{\kap}_1+3\bar{\lam}_0^2-2\alp \bar{\lam}_0\bar{\lam}_1-3\alp\bar{\lam}_0\bar{\kap}_1 +\left(-\alp+5-\frac{\pi^2}{2} \right)\bar{\kap}_0\bar{\kap}_1
\right. \\
  & \qquad \left. {}
+\left( \frac{\alp^2}{2}+1-\frac{\pi^2}{8}\right)\bar{\kap}_1^2+\left( \frac{\alp^2}{2}-\frac{\pi^2}{24} \right)\bar{\kap}_1\bar{\kap}_2  \right] .
\end{split}
\label{ratio-JF}
\ee

Now, taking advantage of some relations that associate the HSRPs in the two frames (see~\cite{Karam2017})
we showed that the inflationary parameters coincide
\begin{eqnarray}
\hat{n}_S = \bar{n}_S , \\
\hat{r} = \bar{r} \,.
\end{eqnarray}
and therefore the frames are equivalent. Since the Green's function method is valid up to arbitrary order in the slow-roll expansion, we expect the equivalence between the spectral indices in the Jordan and Einstein frames to extend to all orders.

Finally, by using the third-order Taylor expansions of the HSRPs in terms of the potential ones (see~\cite{Karam2017} for the full expressions) we can express the inflationary indices in terms of the latter, which are manifestly invariant.
\be 
\begin{split}
n_{s} =&1
 -6 \eps_V+2\eta_V+\left(24 \alp -\frac{10}{3} \right)\eps^2_V-\left( 16 \alp+2\right)\eps_V \eta_V+\frac{2}{3} \eta_V^2+\left( 2 \alp+\frac{2}{3}\right) \zeta_V^2
\\
& -\left(90 \alp^2-\frac{104}{3}\alp+\frac{3734}{9}-\frac{87 \pi^2}{2} \right)\eps_V^3+\left( 90 \alp^2+\frac{4}{3} \alp+\frac{1190}{3}-\frac{87\pi^2}{2} \right) \eps_V^2 \eta_V
\\
& 
- \left( 16 \alp^2+12\alp+\frac{742}{9}-\frac{28 \pi^2}{3}\right) \eps_V \eta_V^2-\left(12 \alp^2+4 \alp+\frac{98}{3}-4 \pi^2 \right)\eps_V \zeta_V^2
\\
& 
+\left(\alp^2+\frac{8}{3}\alp+\frac{28}{3}-\frac{13 \pi^2}{2} \right) \eta_V \zeta^2_V+\frac{4}{9}\eta_V^3+\left(\alp^2+\frac{2}{3}\alp+\frac{2}{9}-\frac{\pi^2}{12} \right) \rho_V^3 ,
\end{split}
\label{scalar-index-V}
\ee
\be 
\begin{split}
r =&
16 \eps_V \left[1-\left(4 \alp+\frac{4}{3} \right) \eps_V+\left( 2\alp +\frac{2}{3}\right) \eta_V+\left(16 \alp^2+\frac{28}{3}\alp+\frac{356}{9}-\frac{14 \pi^2}{3} \right) \eps_V^2
\right. \\
  &\quad \left. {}
-\left(14 \alp^2+10 \alp+\frac{88}{3}-\frac{7\pi^2}{2} \right)\eps_V \eta_V+\left(2 \alp^2+2 \alp+\frac{41}{9}-\frac{\pi^2}{2} \right) \eta_V^2
\right. \\
  &\quad \left. {}
+\left(\alp^2+\frac{2}{3}\alp+\frac{2}{9}-\frac{\pi^2}{12} \right)\zeta_V^2 \right]
\end{split}
\label{ratio-V}
\ee
We thus have fully invariant results. In a given model, once we derive the invariant potential $\Iv$ in terms of the invariant field $\Iphi$, we can readily obtain the PSRPs and express the inflationary observables in an invariant way in terms of $\Iv$ and its derivatives. 

\section{Number of $e$-folds}

Nevertheless, since the time parameter is different in the two frames, we have different definitions for the number of $e$-folds, which differ by a factor that depends on the first invariant
\begin{equation}
\mathrm{d} \bar{N} = \mathrm{d} \hat{N} + \frac{1}{2} \, \mathrm{d} \ln \Inm = \left( - \frac{1}{\sqrt{\epsilon_H}} + \frac{1}{2} \frac{\mathrm{d} \ln \Inm}{\mathrm{d} \Iphi} \right) \mathrm{d} \Iphi
\end{equation}
In order to quantify this difference, let us consider as an example a non-minimal Coleman--Weinberg inflationary model~\cite{Kannike2016}\footnote{See also~\cite{Kannike2017a, Artymowski2017, Racioppi2017, Karam2018a, Racioppi2018, Antoniadis2018, Karam2019} for more considerations on this model.}, specified by the model functions
\begin{eqnarray}
\mathcal{A} (\Phi) &=& \xi \Phi^2 \,, \quad \mathcal{B} (\Phi) = 1 \,, \quad \sigma(\Phi) = 0 \\
\mathcal{V} (\Phi) &=& \Lambda^4 + \frac{1}{8} \beta_{\lambda_\Phi} \left( \ln \frac{\Phi^2}{v^2_\Phi} - \frac{1}{2} \right) \Phi^4 \,, \quad 1 = \xi v^2_\Phi
\end{eqnarray}
where the cosmological constant $\Lambda^4$ was included in order to realize $\mathcal{V} (v_\Phi) = 0$ and $\beta_{\lambda_\Phi}$ is the beta function of the quartic scalar coupling $\lambda_\Phi$. Furthermore, in this model the Planck scale is dynamically generated through the VEV of the scalar field $v_\Phi$. Upon minimizing the effective potential one finds
\begin{equation}
\mathcal{V} (\Phi) = \Lambda^4 \left\lbrace 1 + \left[ 2 \ln \left( \frac{\Phi^2}{v^2_\Phi} \right) - 1 \right] \frac{\Phi^4}{v^4_\Phi}   \right\rbrace \,, \qquad  \Iphi = \sqrt{\frac{1 + 6 \xi}{2 \xi}} \ln \left( \frac{\Phi}{v_\Phi} \right)
\end{equation}
Then, the invariant potential $\Iv$ in terms of $\Iphi$ takes the form
\begin{equation}
\Iv = \Lambda^4 \left( 4 \sqrt{\frac{2 \xi}{1 + 6 \xi}} \, \Iphi +  e^{- 4 \sqrt{\frac{2 \xi}{1 + 6 \xi}} \Iphi }  - 1 \right)
\end{equation}
and is interesting because for small $\xi$ it approaches quadratic inflation (which is excluded),
\begin{equation}
\Iv \vert_{\xi \rightarrow 0} \sim 16 \, \xi \, \Lambda^4 \, \Iphi^2
\end{equation}
while for larger $\xi$ (above $\sim 0.1$) it goes to the limit of linear inflation (which was still marginally consistent with the Planck 2015 data)
\begin{equation}
\Iv \rvert_{\xi \rightarrow \infty} \sim \frac{4}{\sqrt{3}} \, \Lambda^4 \, \Iphi
\end{equation}

In table~\ref{table:E-vs-J-efolds} we show the predictions we obtain for $n_s$ and $r$ to first and third order for the same number of $e$-folds in the two frames and for various values of the non-minimal coupling $\xi$.
\begin{table}
\begin{center}
\begin{tabular}{| c | c | c | c | c | c |}
\hline
 & & & & & \\[-1em]
 & $n_S^{(\rm I)}$ & $n_S^{(\rm III)}$ & $r^{(\rm I)}$ & $r^{(\rm III)}$ & $\xi$ \\
\hline 
 & & & & & \\[-1em] 
$\hat{N} = 60$ & $0.96702$ & $0.96712$ & $0.12782$ & $0.12552$ & $10^{-5}$ \\
\hline
 & & & & & \\[-1em]
$\bar{N} = 60$ & $0.96699$ & $0.96709$ & $0.12792$ & $0.12562$ & $10^{-5}$ \\
\hline
 & & & & & \\[-1em]
$\hat{N} = 60$ & $0.96935$ & $0.96956$ & $0.09655$ & $0.09466$ & $10^{-3}$ \\
\hline 
 & & & & & \\[-1em]
$\bar{N} = 60$ & $0.96911$ & $0.96933$ & $0.09736$ & $0.09544$ & $10^{-3}$ \\
\hline 
 & & & & & \\[-1em]
$\hat{N} = 60$ & $0.97451$ & $0.97477$ & $0.06796$ & $0.06675$ & $0.1$ \\
\hline 
 & & & & & \\[-1em]
$\bar{N} = 60$ & $0.97320$ & $0.97348$ & $0.07148$ & $0.07013$ & $0.1$ \\
\hline 
 & & & & & \\[-1em]
$\hat{N} = 60$ & $0.97482$ & $0.97507$ & $0.06716$ & $0.06597$ & $10$ \\
\hline
 & & & & & \\[-1em]
$\bar{N} = 60$ & $0.97276$ & $0.97305$ & $0.07264$ & $0.07125$ & $10$ \\
\hline
\end{tabular}
\caption{First and third order results for the observables of the nonminimal Coleman-Weinberg model considered in~\cite{Kannike2016} for various values of the nonminimal coupling $\xi$ and for $\hat{N} = \bar{N} = 60$.}
\label{table:E-vs-J-efolds}
\end{center}
\end{table}
The $n_s$ do not differ much, but $r$ for large $\xi$ differs about $2\%$ from first to third order and about $8\%$ between Einstein and Jordan $e$-folds. These differences might seem small but they are very important in view of the expected sensitivity of future experiments~\cite{Wu2016, Matsumura2013, Aguirre2019} which will measure $r$ with an accuracy of $0.001$\footnote{See also~\cite{Wolfson2018}.}. Finally, for the same invariant field excursion, we find a difference of about $4.5$ $e$-folds for large $\xi$.
\begin{figure}
\centering
\includegraphics[width=11.5cm]{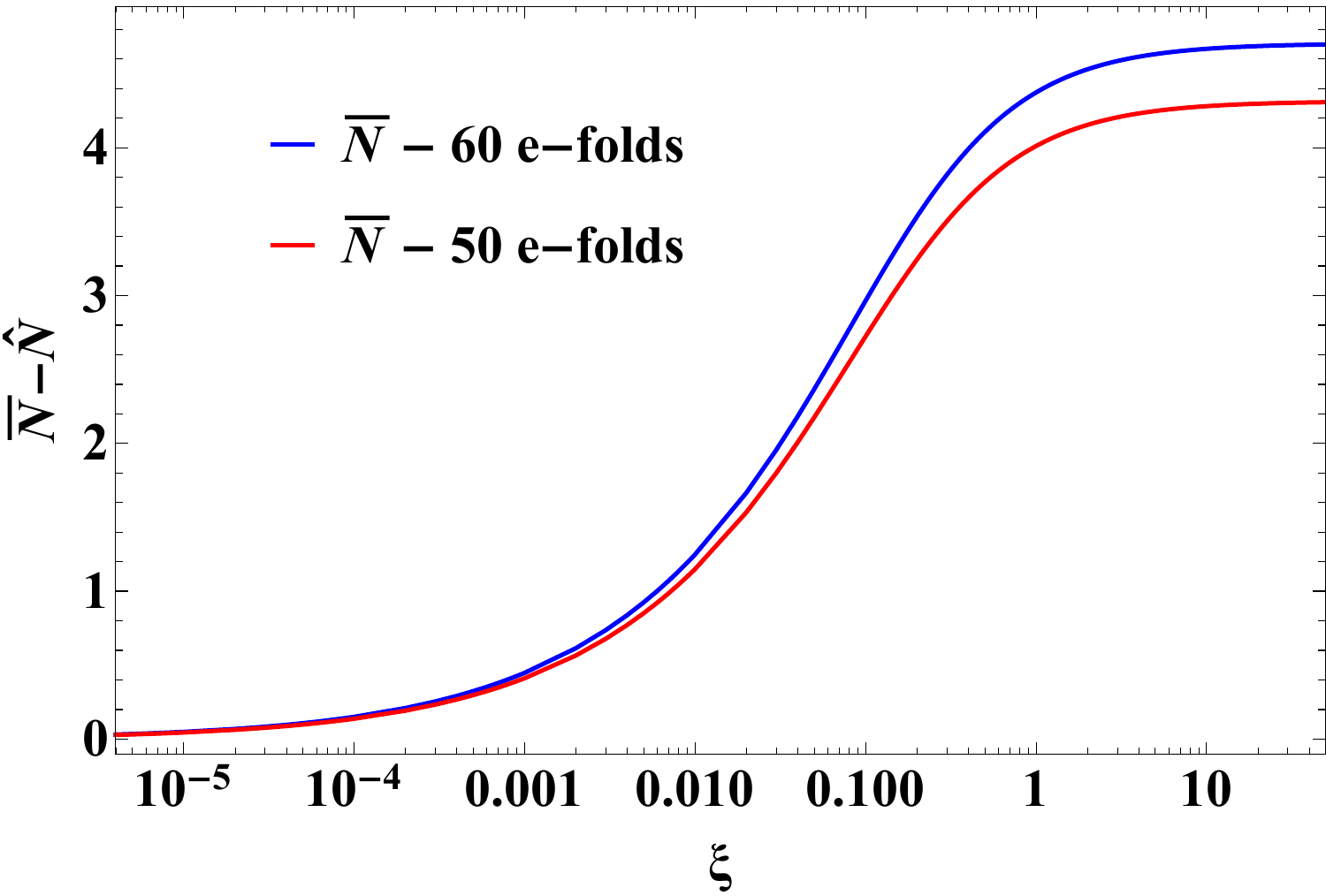}
\caption{The difference between the JF ($\bar{N}$) and the EF ($\hat{N}$) number of $e$-folds as a function of the nonminimal coupling $\xi$ for $\hat{N} = 60$ (top curve) and $\hat{N} = 50$ (bottom curve). We see that as $\xi$ grows we need more $e$-folds in the Jordan frame for the same inflaton field excursion.}
\label{fig:DeltaNFinal}
\end{figure}

\section{Conclusions}

The invariant formalism has been employed in the computation of various cosmological observables~\cite{Jaerv2015, Jaerv2015a, Hohmann2017} and has been generalized to multiscalar-tensor theories~\cite{Kuusk2016a}, scalar-tensor gravity in the Palatini approach~\cite{Kozak2018}, scalar-torsion gravity~\cite{Hohmann2018} and to higher-dimensional scalar-tensor gravity~\cite{Karam2018}. 

In conclusion, we saw that non-minimally coupled theories support viable inflationary models. These models belong to the class of scalar-tensor theories where the frame issue arises. In order to circumvent it we used quantities that are invariant under a conformal rescaling of the metric and a scalar field redefinition. We computed the inflationary observables up to third order in the slow-roll approximation in both frames and showed that they are equivalent. Then, we expressed the observables in terms of the PSRPs which are manifestly invariant. Nevertheless, the definition of the number of $e$-folds differs in the two frames and its effect on the predictions has also been investigated. We regard the Jordan frame definition as the correct one since it takes into account all three basic invariants and also includes the Einstein frame definition.

\section*{Acknowledgments}
A.K. and T.P. acknowledge support from the Operational Program ``Human Resources Development, Education and Lifelong Learning" which is co-financed by the European Union (European Social Fund) and Greek national funds. 
A.K. also thanks the organizers of the “Workshop on the Standard Model and Beyond", Corfu 2018, for the hospitality during his stay, the financial support, and for giving him the opportunity to present this work.

\bibliography{References}{}
\bibliographystyle{utphys}

%

\end{document}

%% file: macros.tex
\newcommand{\newc}{\newcommand*}

\long\def\begincomment#1\endcomment{%
        \begingroup\sf\baselineskip12pt#1\endgroup}

\newc{\etal}{\textrm{et al.}} 
\newc{\eg}{\textrm{e.g.}} 
\newc{\ie}{\textrm{i.e.}}
\newc{\etc}{\textrm{etc}}
\newc\vs{\textrm{vs.}}
\newc{\cl}{\rm {C.L.}}

\newc{\eV}{\ensuremath{\,\mathrm{eV}}}
\newc{\KeV}{\ensuremath{\,\mathrm{keV}}}
\newc{\MeV}{\ensuremath{\,\mathrm{MeV}}}
\newc{\GeV}{\ensuremath{\,\mathrm{GeV}}}
\newc{\TeV}{\ensuremath{\,\mathrm{TeV}}}

\newc{\invpb}{\ensuremath{/\text{pb}}}
\newc{\invfb}{\ensuremath{\,\text{fb}^{-1}}}
\newc\nb{\ensuremath{\,\mathrm{nb}}} \newc\pb{\ensuremath{\,\mathrm{pb}}} \newc\fb{\ensuremath{\,\mathrm{fb}}}
\newc\pc{\ensuremath{\,\mathrm{pc}}}
\newc\kpc{\ensuremath{\,\mathrm{kpc}}}
\newc\mpc{\ensuremath{\,\mathrm{Mpc}}}
\newc\ps{\ensuremath{\,\mathrm{ps}}} 
\newc\cmeter{\ensuremath{\,\mathrm{cm}}} 
\newc\meter{\ensuremath{\,\mathrm{m}}} 
\newc\kmeter{\ensuremath{\,\mathrm{km}}}
\newc\second{\ensuremath{\,\mathrm{s}}}
\newc\msecond{\ensuremath{\,\mathrm{ms}}}
\newc\nsecond{\ensuremath{\,\mathrm{ns}}}
\newc\psecond{\ensuremath{\,\mathrm{ps}}}

\newc{\D}{\partial}
\newc{\shh}{\lambda_{h}}
\newc{\shf}{\lambda_{h\phi}}
\newc{\sff}{\lambda_{\phi}}
\newc{\sss}{\lambda_{s}}
\newc{\ssisi}{\lambda_{\sigma}}
\newc{\sssi}{\lambda_{s \sigma}}
\newc{\shs}{\lambda_{h s}}
\newc{\shsi}{\lambda_{h \sigma}}
\newc{\sfs}{\lambda_{\phi s}}
\newc{\sfsi}{\lambda_{\phi \sigma}}

\newcommand{\dd}{\mathrm{d}}

\newcommand{\Inm}{\mathcal{I}_m}
\newcommand{\Iv}{\mathcal{I}_{\mathcal{V}}}
\newcommand{\Iphi}{\mathcal{I}_\phi}
\newcommand{\hatt}{\hat{t}}
\newcommand{\bart}{\bar{t}}
\newcommand{\hk}{\hat{\kappa}}
\newcommand{\bk}{\bar{\kappa}}

\newcommand{\bl}{\bar{\lambda}}
\newcommand{\heps}{\hat{\epsilon}}

\newcommand{\hH}{\hat{H}}
\newcommand{\bH}{\bar{H}}

\let\eps=\epsilon
\let\lam=\lambda
\let\kap=\kappa
\let\alp=\alpha

\newc{\chisqmin}{\ensuremath{\chi^2_{\mathrm{min}}}}
\newc{\Delchisq}{\ensuremath{\Delta\chi^2}}
\newc{\chisq}{\ensuremath{\chi^2}}
\newc{\like}{\ensuremath{\mathcal{L}}}

\newc\lsim{\ensuremath{\mathrel{\rlap{\lower4pt\hbox{\hskip1pt$\sim$}}\raise1pt\hbox{$<$}}}}
\newc\gsim{\ensuremath{\mathrel{\rlap{\lower4pt\hbox{\hskip1pt$\sim$}}\raise1pt\hbox{$>$}}}}
\newc{\VEV}[1]{\ensuremath{\langle #1 \rangle}}
\newc{\dl}{\ensuremath{\stackrel{\leftarrow}{D}}}
\newc{\dr}{\ensuremath{\stackrel{\rightarrow}{D}}}

\newc{\bcenter}{\begin{center}}    \newc{\ecenter}{\end{center}}
\newc{\bfl}{\begin{flushleft}}    \newc{\efl}{\end{flushleft}}
\newc{\bfr}{\begin{flushright}}    \newc{\efr}{\end{flushright}}

\newc{\bi}{\begin{itemize}}
\newc{\ei}{\end{itemize}}
\newc{\bed}{\begin{description}}
\newc{\eed}{\end{description}}
\newc{\ben}{\begin{enumerate}}
\newc{\een}{\end{enumerate}}

\newc{\be}{\begin{equation}}
\newc{\ee}{\end{equation}}
\newc{\ba}{\begin{eqnarray}}
\newc{\ea}{\end{eqnarray}}
\newc{\ra}{\rightarrow}
\def\beq{\begin{equation}}
\def\eeq{\end{equation}}
\def\bea{\begin{eqnarray}}
\def\eea{\end{eqnarray}}
\def\bs#1\es{\begin{split}#1\end{split}}
\def\bsub#1\esub{\begin{subequations}#1\end{subequations}}

\newc{\alphas}{\ensuremath{\alpha_s}}
\newc{\alphatwo}{\ensuremath{\alpha_2}}
\newc{\alphaone}{\ensuremath{\alpha_1}}
\newc{\alphai}[1]{\ensuremath{\alpha_{#1}}}
\newc{\alphaem}{\ensuremath{\alpha_{\mathrm{em}}}}
\newc{\alphaeff}{\ensuremath{\alpha_{\mathrm{eff}}}}
\newc{\sineff}{\ensuremath{\sin \theta_{\mathrm{eff}}}}
\newc{\sinsqeff}{\ensuremath{\sin^2 \theta_{\mathrm{eff}}}}
\newc{\dalphahad}{\ensuremath{\Delta \alpha_{\mathrm{had}}}}
\newc{\yt}{\ensuremath{h_t}} \newc{\yb}{\ensuremath{h_b}} \newc{\ytau}{\ensuremath{h_{\tau}}}
\newc\mz{\ensuremath{M_Z}} 
\newc\mw{\ensuremath{m_W}}
\newc\mZ{\mz}        \newc\mW{\mw}
\newc\mhsm{\ensuremath{ m_{H_{\mathrm{SM}}}}}
\newc{\mtop}{\ensuremath{ m_t}}               \newc{\mtpole}{\ensuremath{ M_t}}
\newc{\mbottom}{\ensuremath{ m_b}} 
\newc{\mtau}{\ensuremath{ m_{\tau}}}
\newc{\mt}{\mtpole}
\newc{\mb}{\mbottom} 
\newc{\rgg}{\ensuremath{R_{h}(\gamma\gamma)}}
\newc{\rzz}{\ensuremath{R_{h}(ZZ)}}
\newc{\rtwogg}{\ensuremath{R_{h_2}(\gamma\gamma)}}
\newc{\rtwozz}{\ensuremath{R_{h_2}(ZZ)}}
\newc{\ronegg}{\ensuremath{R_{h_1}(\gamma\gamma)}}
\newc{\ronezz}{\ensuremath{R_{h_1}(ZZ)}}
\newc{\rsiggg}{\ensuremath{R_{h_\textrm{sig}}(\gamma\gamma)}}
\newc{\rsigzz}{\ensuremath{R_{h_\textrm{sig}}(ZZ)}}
\newc{\llbar}{\ensuremath{\ell\bar{\ell}}}
\newc{\tauptaum}{\ensuremath{ \tau^+\tau^-}}
\newc{\qqbar}{\ensuremath{ q\bar{q}}} \newc{\ppbar}{\ensuremath{ p\bar{p}}}
\newc{\bbbar}{\ensuremath{ b\bar{b}}} \newc{\ttbar}{\ensuremath{ t\bar{t}}}
\newc{\ffbar}{\ensuremath{ f\bar{f}}} \newc{\tautaubar}{\ensuremath{ \tau\bar{\tau}}}
\newc{\mchi}{\ensuremath{m_{\chi}}}
\newc{\squark}{\ensuremath{\tilde{q}}}
\newc{\slepton}{\ensuremath{\tilde{l}}}
\newc{\gluino}{\ensuremath{\tilde{g}}} 
\newc{\wino}{\ensuremath{\tilde{W}}}
\newc{\bino}{\ensuremath{\tilde{B}}}
\newc{\mgluino}{\ensuremath{{m_{\gluino}}}}
\newc{\tone}{\ensuremath{{\tilde{t}_1}}}